\documentclass[%
 preprint,
nofootinbib,
 amsmath,amssymb,
 aps,
prd,
]{revtex4-1}
\usepackage{natbib}	
\usepackage{color}
\usepackage{mathtools}
\usepackage{amssymb}
\usepackage{amsfonts}
\usepackage{latexsym}
\usepackage{graphicx}
\usepackage{dcolumn}
\usepackage{bm}

\newcommand{\de}{{de}}
\newcommand{\s}{{\rm s}}

\newcommand{\lam}{\lambda}

\newcommand{\cD}{{\cal D}}

\newcommand{\gam}{\gamma}

\newcommand{\ve}{\varepsilon}
\newcommand{\al}{\alpha}
\newcommand{\vphi}{\varphi}

\newcommand{\bee}{\begin{equation}}
\newcommand{\een}{\end{equation}}
\newcommand{\Gam}{\Gamma}

\def\bea{\begin{eqnarray}}
\def\eea{\end{eqnarray}}
\def\ba{\begin{eqnarray}}
\def\ea{\end{eqnarray}}
\def\p{\partial}
\def\2{\sqrt{2}}

\def\mc{\mathcal}

\def\be{\begin{equation}}
\def\ee{\end{equation}}
\def\bea{\begin{eqnarray}}
\def\eea{\end{eqnarray}}
\def\a{\alpha}
\def\b{\beta}
\def\g{\gamma}
\def\G{\Gamma}
\def\de{\delta}

\def\l{\lambda}

\def\k{\kappa}
\def\m{\mu}
\def\n{\nu}
\def\r{\rho}

\def\s{\sigma}

\def\t{\tau}

\def\vf{\varphi}

\def\d{\nabla}
\def\tG{\tilde{\Gamma}}
\def\gam{\gamma}
\def\tL{\tilde{L}}
\def\tQ{\tilde{Q}}
\def\tP{\tilde{P}}

\begin{document}


\title{Covariantised vector Galileons}

\author{Matthew Hull}
 \email{matthew.hull@port.ac.uk}
 \author{Kazuya Koyama}%
 \email{kazuya.koyama@port.ac.uk}
\affiliation{Institute of Cosmology \& Gravitation, University of Portsmouth, Dennis Sciama Building, Portsmouth, PO1 3FX, United Kingdom}


\author{Gianmassimo Tasinato}
\email{g.tasinato@swansea.ac.uk}
\affiliation{Department of Physics, Swansea University, Swansea, SA2 8PP, U.K.}


\date{\today}
\begin{abstract}
Vector Galileons are ghost-free  systems containing
higher derivative interactions of vector fields. They  break  the vector gauge symmetry, and the dynamics
 of the longitudinal vector polarizations acquire a Galileon symmetry in an appropriate decoupling limit in Minkowski space.
 Using an ADM approach, we carefully reconsider the coupling with gravity 
of vector Galileons, with the aim of studying the necessary conditions to avoid the propagation of ghosts. We develop arguments that put on a more solid footing  the results previously obtained in the literature. Moreover, working in analogy with the 
scalar counterpart, we find indications for the existence
of a `beyond Horndeski' theory involving vector degrees of freedom, that avoids the propagation of ghosts
thanks to secondary constraints. In addition, we analyze  a Higgs mechanism
for generating vector Galileons through  spontaneous symmetry breaking, and we present its consistent covariantisation.
\end{abstract}

\maketitle

\section{\label{sec:intro} Introduction}
Models of infrared modifications of gravity have been the focus of intense study in cosmology for over a decade.\footnote{For a comprehensive review see \cite{Clifton:2011jh, Koyama:2015vza}.}  Driven by the desire to find alternative explanations for the observed acceleration in the expansion of the universe \cite{Riess:1998cb, Perlmutter:1998np}, model builders have mostly concentrated on using scalar fields for constructing candidate models of dark energy or modified gravity.  Furthermore, due to strict solar system constraints \cite{Bertotti:2003rm}, any modified theory of gravity worthy of playing a role in the expansion of the universe, needs to approach general relativity on small scales. \emph{Screening mechanisms} are a realization of this behavior and can originate from non-linearities in either the scalar's potential or its kinetic term.  A particularly robust mechanism is found within a set of theories that have non-linear derivative self-interactions.  This mechanism, called the \emph{Vainshtein mechanism} was first realized in massive gravity \cite{Vainshtein:1972sx} and it was rediscovered in investigations of specific braneworld scenarios \cite{Dvali:2002pe,Luty:2003vm} (see \cite{Babichev:2013usa} for a review). An analysis of the decoupling limit of the DGP braneworld theory led to the discovery of scalar fields, called \emph{Galileons}, with the Galilean field redefinition symmetry: $\phi\to\phi+b_{\m}x^{\m}+c$ \cite{Nicolis:2008in}.\footnote{See also \cite{Gabadadze:2012tr} for a generalisation to bulks with different isometries.}  Issues concerned with the stability of these theories in a dynamical space-time led to the formulation of a non-minimal covariantisation scheme \cite{Deffayet:2009wt}; the later generalization of which, recovered the Horndeski action \cite{Deffayet:2009mn,Horndeski:1974wa}.  Most recently, it has been discovered that, contrary to general expectation, the minimal covariantisation of the Galileons is in fact ghost free \cite{Gleyzes:2014dya,Lin:2014jga,Gleyzes:2014qga,Gao:2014soa}.  This is due to the fact that even though the equations of motion contain derivatives of third order, there exists a \emph{hidden} second order constraint equation which allows one to replace the third order time derivatives with lower order expressions \cite{Deffayet:2015qwa}. \\
In this work we will discuss the viability of using vector fields as an alternative candidate for dark energy.  Given that vector fields are  able to communicate long range forces, it is natural to ask whether the special properties of the scalar field models can also be realized for vectors.  In fact, the special non-linear structure of the Galileon theories has already been extended to general p-forms \cite{Deffayet:2010zh}, including a version for the gauge field strength tensors.  Here, we follow the work of \cite{Tasinato:2014eka,Heisenberg:2014rta} and abandon gauge symmetry by directly endowing the vector fields with non-linear derivative self-interactions.  These vector fields, dubbed here as \emph{vector Galileons} can be seen as a non-linear extension of Proca theory and have been shown to have interesting cosmological applications \cite{Tasinato:2014eka,Tasinato:2014mia} associated with the dynamics of the vector longitudinal polarization.  Indeed, in \cite{Tasinato:2014eka} a Horndeski inspired non-minimal covariantisation for the quartic was found to generate a technically natural effective cosmological constant.  Furthermore, including an additional large bare cosmological constant $\Lambda_{bare}$ modifies the Friedman equation with a term proportional to the inverse of $\Lambda_{bare}$.  This fact, together with the existence of technically natural parameters, could provide a new opportunity to resolve the `Old Cosmological Problem' \cite{Tasinato:2014mia}.
As a consequence of the relationship with Proca theory, the phenomenology of these models is further enhanced by a corresponding extension to the Higgs mechanism that generates vector Galileons dynamically \cite{Hull:2014bga} (see \cite{Kamada:2010qe} for an earlier model developed in the context of inflation).\\
Of course, models of vector dark energy have a long history: see e.g., \cite{Koivisto:2008xf}, or the discussion in the review \cite{Clifton:2011jh}. The advantage of our formulation is the connection with scalar Galileons \cite{Nicolis:2008in}. Indeed, as shown in  \cite{Tasinato:2014eka},  in an appropriate decoupling limit the vector longitudinal mode
decouples from the transverse vector polarizations and acquires scalar Galileon self-interactions. This nice feature
can have important consequences for the stability of this theory under quantum corrections, given
the powerful non-renormalization theorems of Galileon theories \cite{Luty:2003vm,Nicolis:2004qq}.\\

The aim of this work is to study the coupling to gravity of these derivatively coupled vector models.
As with the scalar-tensor theories, it is not immediately clear whether vector Galileons are ghost free around a dynamical space-time.  However, by analogy with the scalar case, a covariantised system was suggested in \cite{Tasinato:2014eka}, and a generalized \emph{vector-Horndeski} system was suggested in \cite{Heisenberg:2014rta}.\\  
In section \ref{sec:cvg} we discuss the construction of this theory and comment on the evidence we find for its mathematical consistency.  In section \ref{ssec:ms} we follow the example given for $G^3$ \emph{beyond Horndeski} theories \cite{Gleyzes:2014dya} and make use of a special ansatz for the vector field to find similar evidence for the consistency of the minimal covariantisation of vector Galileons -- i.e., a scenario in which partial derivatives are substituted with covariant derivatives of the vector, with no further couplings to the metric.
   Lastly, in section \ref{Sec:CG-Higgs} we present a non-minimal covariantisation of the Galileonic Higgs model from \cite{Hull:2014bga}.\footnote{The covariantisation of similar gauged Galileon theories were also studied in \cite{Zhou:2011ix,Goon:2012mu}.}  This is achieved by satisfying the dual requirements of both stability and $U(1)$ invariance.  We conclude in section \ref{Sec:Conc}.

\paragraph{Notation}
We make use of the \textit{Levi-Civita epsilon tensor} throughout the text.  In particular, we make use of the following property 
\bee
	\ve_{\gam_1 \dots \gam_{D-n} \al_{1}\dots \al_{n}}\ve^{\gam_1 \dots \gam_{D-n} \beta_{1}\dots \beta_n} = -(D-n)!\,n!\,\delta^{[\beta_{1}\dots \beta_n]}_{\al_{1}\dots\al_n}\,.
\een
We will find it convenient to make the following definition, $\pi_{\mu_1\dots\mu_n} \equiv \p_{\mu_n}\dots \p_{\mu_1}\pi$.
Four dimensional indices are written with greek lower case letters: $\mu,\nu,\ldots$ whereas for the three dimensional indices we use lower case latin: $i,j,k\ldots$.  The three-metric is written $\gamma_{ij}$ and raises and lowers three dimensional objects like the extrinsic curvature, $K_{ij}$, or the three dimensional Riemann tensor $\tilde{R}_{ijkl}$.  The four dimensional covariant derivative is written as $\d_{\mu}$ and the three dimensional covariant derivative, (which is compatible with $\gamma_{ij}$) is written as $D_{i}$.  The corresponding four dimensional connection is written as $\Gamma^{\mu}_{\nu\rho}$ and the three dimensional one as $\tilde{\Gamma}^{i}_{jk}$.

\paragraph{ADM decomposition}
We will make extensive use of the ADM formalism and in particular our discussion will refer to the lapse, $N$ and extrinsic curvature, $K_{ij}$ which are properly presented in appendix \ref{sec:ADM} but which we define here for convenience.\\
Given a foliation of a four dimensional space-time by three surfaces, the metric, $g_{\mu\nu}$, can be decomposed in terms of components normal and tangent to the surfaces as
\bee
	g_{\mu\nu} = -\big( N^2-N_iN^i\big)dt^2 +2 N_i dt dx^i + \gamma_{ij}dx^i dx^j\,,
\een
where $N$ is called the \emph{lapse} and $N_i$ is the \emph{shift}.  The extrinsic curvature of the foliating three surfaces is defined as
\bee\label{eq:Kintro}
	K_{ij}\equiv - \d_{i}n_{j} = \Gamma^{\mu}_{ij}n_{\mu} = -N \Gamma^{0}_{ij} = \frac{1}{2N}\Big( D_{i}N_{j}+D_{j}N_{i}- \dot{\gam}_{ij} \Big)\,.
\een

\section{ \label{sec:cvg} Covariantisation of vector Galileons}

Vector Galileons are ghost-free
 systems containing derivative self-interactions of vector fields, 
 that break gauge symmetry and can have interesting cosmological consequences.   
As explained in the introduction, the aim of this paper is to use an ADM approach
to  reconsider more carefully the consistency of the covariant couplings
of vector Galileons with gravity, that were first introduced and studied in \cite{Tasinato:2014eka,Heisenberg:2014rta} by making use of the analogy with the scalar Galileon counterparts. We start by discussing non-minimal couplings of vector Galileons
 with gravity, studying the conditions to avoid the propagation of ghosts. We continue by discussing minimal couplings
 with gravity, with the purpose of investigating possible  vector-tensor counterparts to the beyond Horndeski scalar-tensor theories of \cite{Gleyzes:2014dya}. 
\subsection{\label{ssec:cvGNMC} Covariantisation of vector Galileons via the use of non-minimal couplings} 
As in the case for scalar Galileons, there are different forms for the vector Galileons which are related by a total derivative.  In addition to this, the vectors also have two extra free parameters, $\big(c_2, d_3\big)$, due to the ability to generate ghost free terms of the form, $f_{2}(A^2, A\cdot F, F^2, FF^*)$\cite{Heisenberg:2014rta,Gripaios:2004ms}.\footnote{ We follow \cite{Heisenberg:2014rta} and use $A \cdot F$ to denote all possible contractions of $A_{\mu}$ with $F_{\mu \nu}$ and $\big(c_2, d_3\big)$ to denote the two extra parameters.  There is some degeneracy here, for example, starting with the quartic, $\mc{L}^{(4)}_{vG}$, we can use integrations by parts to find expressions like $A^2F^2$ and $A_{\mu}A_{\nu}F^{\mu\r}F^{\nu}_{\r}$.}\\
We use the antisymmetric properties of the \textit{Levi-Civita epsilon tensor} to write the vector Galileons on Minkowski space-time as
\begin{align} \label{eq:vGsys1}
	\mc{L}_{F} &= -\frac{1}{4}F_{\mu\nu}F^{\mu\nu}\,,\\
	\mc{L}^{(2)}_{vG} &= A_{\mu}A^{\mu}\,,\\
	\mc{L}^{(3)}_{vG} & = \frac{1}{2}\,\ve^{\m_{1} \mu_{3} \l \s}\ve^{\n_{2} \nu_{4} }_{\quad \,\,\,\,\l \s} \,A_{\m_{1}}A_{\n_{2}}\, A_{\mu_{3}\nu_{4}}\,,\\
	\mc{L}^{(4)}_{vG} & = \ve^{\m_{1} \mu_{3}\mu_{5} \l}\ve^{\n_{2} \nu_{4} \nu_{6}}_{\quad \quad\,\, \l} \,A_{\m_{1}}A_{\n_{2}}\, \big(A_{\mu_{3}\nu_{4}}A_{\mu_{5}\nu_{6}}+c_{2}F_{\m_{3}\m_{5}}F_{\n_{4}\n_{6}}\big)\label{eq:vGsys4}\,,\\
	\mc{L}^{(5)}_{vG} & = \ve^{\m_{1} \mu_{3}\mu_{5} \mu_{7}}\ve^{\n_{2} \nu_{4} \nu_{6}\nu_{8}} \,A_{\m_{1}}A_{\n_{2}}\, \big(A_{\mu_{3}\nu_{4}}A_{\mu_{5}\nu_{6}}A_{\mu_{7}\nu_{8}}+d_{3}A_{\m_{3}\n_{4}}F_{\m_{5}\m_{7}}F_{\n_{6}\n_{8}}\big)\,, \label{eq:vGsys2}
\end{align}
where $A_{\mu\nu}\equiv \p_{(\mu}A_{\nu)}$, and $F_{\mu\nu} \equiv \p_{[\mu}A_{\nu]}$,\footnote{Note we will also use this notation to denote the same symmetric/antisymmetric combinations with covariant derivatives.} are respectively normalized symmetric and antisymmetric combinations.  The extra bi-parameter freedom also extends to the covariantisation of these theories. Furthermore, the existence of various forms for the vector Galileon raises questions about whether there is any freedom to choose the form of any additional non-minimal couplings.\\
In this section we address this problem by identifying the potentially unstable terms that are generated when we naively covariantise the derivative interactions in the vector Galileon system.  Indeed, the fact that the system is able to avoid producing Ostrogradsky ghosts relies on the fact that partial derivatives commute together with the antisymmetric sum over the indices.  Minimal covariantisation of the derivatives spoils their, seemingly essential, commutative property and generates extra interaction terms that, a subset of which, appear to be potentially unstable.  It is exactly the need to eliminate these extra terms that fixes the form of the non-minimal coupling.  In this sense we can view the non-minimal couplings as counter terms that cure the theory from unstable gravitational interaction terms.\\
In the following we present the previously proposed `vector-Horndeski' system of \cite{Tasinato:2014eka,Heisenberg:2014rta} and then use the ADM formalism to investigate its consistency by studying the restrictions imposed on its non-minimal couplings of vectors with gravity.

\subsubsection{\label{sec:vH} The vector-Horndeski system}
The non-minimally covariantised vector Galileons which were first presented in \cite{Tasinato:2014eka,Heisenberg:2014rta,Tasinato:2014mia},\footnote{See also \cite{Gripaios:2004ms} for a discussion of an effective field theory for vectors.} can be written in the form resembling the Horndeski system for scalar-tensor theories
\begin{align}
	\mc{L}_{F} &= -\frac{1}{4}\sqrt{-g}F^{\mu\nu}F_{\mu\nu}\,, \\
	\mc{L}^{(2)}_{vH} &= \sqrt{-g} G_{2}(X)\,, \\ 
	\mc{L}^{(3)}_{vH} &= \sqrt{-g} G_{3}(X)A_{\,\,\mu}^{\mu}\,, \\
	\mc{L}^{(4)}_{vH} &= \sqrt{-g}G_{4}(X)R+ \sqrt{-g} G_{4,X}\, \ve^{\mu \r \lam_{1} \lam_{2}} \ve^{\nu \s}_{\quad \lam_{1}\lam_{2}} \big(A_{\mu \nu}A_{\r \s} + c_{2}F_{\mu\r}F_{\nu\s}\big) \,, \\
	\mc{L}^{(5)}_{vH} &= \sqrt{-g} G_{5}(X)A_{\mu\nu}G^{\mu\nu} -\frac{1}{6}G_{5,X}\, \ve^{\mu \r \g\lam} \ve^{\nu \s\k}_{\quad\,\,\, \lam} \big(A_{\mu \nu} A_{\r\s}A_{\g\k} +d_{3}A_{\mu\nu}F_{\r\g}F_{\s\k}\big)\,,
\end{align}
where $X \equiv -\frac{1}{2}A_{\mu}A^{\mu}$ and $G_{N,X}\equiv \frac{\p G_N}{\p X}$.\\
Whereas earlier work motivated this system via its similarity in construction to Horndeski theory, in the following sections, we analyze the consistency of the covariantised model by focussing on the role of the non-minimal couplings.  

\subsubsection{\label{sec:nmcvg4} Non-minimal covariantisation of the quartic vector Galileon} 
In this section we examine how the inclusion of a specific non-minimal coupling term is able to `cure' a potential instability arising from the covariantisation of the derivatives in the quartic vector Galileon, given by equation (\ref{eq: L4}) below.  Although not necessary, we begin by choosing a certain special ansatz for the vector field and find that within the ADM formalism and at the level of the action, this possible instability is related to the existence of terms of the form, $A_{0}\dot{N}$ and $\dot{A}_{0}N$.  Such terms in the action can produce dynamics for the lapse, $N$ or $A_0$ which would be an extra degree of freedom which does not exist in general relativity nor electromagnetism, where the equations of motion for $N$ and $A_0$ appear as a constraint.  Typically, this extra degree of freedom is identified as a propagating ghost.  

Note however that the existence or non-existence of these terms by themselves do not guarantee that this is or isn't a true classical instability.  It was recently shown in \cite{Domenech:2015tca} that the existence of time derivatives of the lapse does not necessarily mean we have a pathology.  Indeed, they showed that one could start with an Einstein-Hilbert action, perform a transformation that results in a theory with time derivatives of the lapse but as long as the transformation is regular and invertible, the number of degrees of freedom remain invariant. In this case, a full Hamiltonian analysis needs to be performed to confirm the number of propagating degrees of freedom. 

Furthermore, it is true that other terms of the form $A_{0}\dot{N}_{i}$ and $\dot{A}_{0}N_{i}$, could also produce classically unstable dynamics, however, if they were to exist, such terms would represent a far more dramatic increase in the number of degrees of freedom of the theory and therefore as a first step we 
 do not consider this possibility. \\ 
We choose an ansatz for the vectors by setting the spatial components of the vector to zero, $A_{\mu} = \big(A_{0},\vec{0}\big)$ and consider the covariantisation of the quartic vector Galileon after an integration by parts which is given by\footnote{This can be identified with $\mc{L}^{(4)}_{vH}$ in section \ref{sec:vH} by setting $G_2 = G_3 =G_5 =0$, $ c_2 = -\frac{1}{2}$, and $G_4 =-X^2$.}  
\bee \label{eq: L4}
	\mc{L}^{(4)}_{vH} =\sqrt{-g}A_{\sigma}A_{\lambda} g^{\sigma \lambda }\Big( \d^{\mu}A_{\mu}\d^{\nu}A_{\nu} - \d^{\mu}A_{\nu}\d^{\nu}A_{\mu} - \frac{1}{4}A_{\mu}A_{\nu}g^{\mu\nu}R\Big)\,. 
\een
Focussing on the derivative terms we find that after cancellations we are left with a term which contains
\begin{align} \label{eq:vG4unstable}
	\mc{L}^{(4)}_{vH|A} =& \,2\sqrt{-g}A_{0}A_{0} g^{00}\Big( \d^{0}A_{0}\d^{i}A_{i} - \d^{0}A_{i}\d^{i}A_{0} \Big)\,, \\
	\supseteq& \,2 \sqrt{\gam} N A_{0}^2 \Big( -\frac{1}{N^2}\Big) \big( \dot{A}_{0} - \frac{\dot{N}}{N}A_{0}\big)\frac{KA_{0}}{N^3}\,, \notag \\ 
	\Rightarrow \mc{L}^{(4)}_{vH|A}  \supseteq& \,- \frac{1}{2}\sqrt{\gam}(A_{0})^{4}\Big( \frac{\dot{K}}{N^4}\Big)\,,
\end{align}
where we use the symbol `$\supseteq$' to denote that $\mc{L}$ contains this expression amongst other terms.  With our ansatz this is the only term originating from the derivative structure to contain potential instabilities of the form $A_{0}\dot{N}$ and $\dot{A}_{0}N$.  For the non-minimal term, we use the results from appendix \ref{sec:UnstCurv} to find that it contributes
\begin{align}
	\mc{L}^{(4)}_{vH|B}= -\sqrt{-g}\frac{1}{4}A^4R = & -\frac{1}{4}N\sqrt{\gam}(A_{0})^{4}\big(g^{00}\big)^{2}R\,, \notag \\ 
	\supseteq&  \frac{1}{2}\sqrt{\gam}(A_{0})^{4}\Big( \frac{\dot{K}}{N^4}\Big)\,,
\end{align}
which cancels the contribution from the previous derivative term.  Interestingly, this contribution comes from what would have been a total divergence in the Einstein-Hilbert action (notice indeed that in this case the Ricci scalar does not stand alone, but  is weighted by the fourth power of the gauge field).  The addition of a non-minimal coupling thus contributes a new derivative term that after integration by parts has the right structure to cancel the derivative of the lapse.\\  
Furthermore, at the level of operators, the relationship between covariant derivatives and curvature forms on a general space-time allows one to use the non-minimal coupling as a gravitational counter term.  From this point of view, the tuning for the functional form of the factors in the Horndeski theory ensures that these particular  unstable operators do not appear in the action.
\subsubsection{Non-minimal covariantisation of the quintic vector Galileon} 
In this section we apply the analysis from section \ref{sec:nmcvg4} to the covariantised quintic vector Galileon given by\footnote{This can be identified with $\mc{L}^{(5)}_{vH}$ in section \ref{sec:vH} by setting $G_2 = G_3 =G_4 =d_3=0$, and $G_5 =6X^2$.}
\begin{align}
	\mc{L}^{(5)}_{vH}&= \mc{L}^{(5}_{vH|A} + \mc{L}^{(5}_{vH|B}\,,\\
	\mc{L}^{(5)}_{vH|A}& = \,\sqrt{-g}A_{\mu} A_{\nu}g^{\mu \nu}g^{\r\s}g^{\l \t} g^{\ve \de} \Big(A_{\r\s}A_{\l \t}A_{\ve \de} - 3A_{\r\s}A_{\l\ve}A_{\t\de} +2A_{\r\de}A_{\s\l}A_{\ve\t}  \Big)\,, \\
	\mc{L}^{(5)}_{vH|B}& =  \frac{6}{4}\,\sqrt{-g}A_{\mu} A_{\nu}g^{\mu \nu}g^{\r\s}g^{\l \t} g^{\ve \de} A_{\r}A_{\s}A_{\l\ve}G_{\t\de}\,.
\end{align}
We again use a special ansatz for which $A_{\m} = (A_{0},\vec{0})$ and look for terms of the form $\dot{A}_{0}N$, $A_{0}\dot{N}$.\\
With this ansatz we have that $A_{00} = \dot{A}_{0} - \G^{0}_{00}A_{0} \supseteq \dot{A}_{0} - \frac{\dot{N}}{N}A_{0}$ and $G^{ij} \supseteq \frac{2}{N}\g^{i[j}\g^{k]l}\dot{K}_{kl}$.  Since we are restricting our focus as in section \ref{sec:nmcvg4}, we only need to consider the factors of these terms. $\mc{L}^{(5)}_{vH|A}$ contributes
\bee
	\mc{L}^{(5)}_{vH|A} \supseteq\, 6 \sqrt{\g} N A_{0}^{2} g^{00}A_{00} \big( g^{00}g^{ij}g^{kl}- 2 g^{ij}g^{0k}g^{0l}\big)\big( A_{i[j}A_{k]l}\big)\,,
\een
but with our ansatz the last factor can be expressed as 
\bee
	A_{i[j}A_{k]l} = \frac{1}{2N^{2}}\big( K_{ij}K_{kl}-K_{ik}K_{jl}\big)A_{0}^{2}\,.
\een
Therefore we see that the contribution from $\mc{L}^{(5)}_{vH|A}$ can be expressed as 
\bee
	\mc{L}^{(5)}_{vH|A} \supseteq \frac{6\sqrt{\g}A_{0}^{4}}{2N^{5}}\big(\dot{A}_{0} - \frac{\dot{N}}{N}A_{0}\big)\big(K^{2}-K_{ij}K^{ij} \big)\,.
\een
We use the fact that in the ADM formalism we can write $G^{00} = \tilde{R} + K^{2}- K_{ij}K^{ij}$, to find that the contribution from $\mc{L}^{(5)}_{vH|B}$ is
\bee
	\mc{L}^{(5)}_{vH|B} \supseteq \frac{6\sqrt{\g}A_{0}^{4}}{8N^{5}}\big(\dot{A}_{0} - \frac{\dot{N}}{N}A_{0}\big)\big(K^{2}-K_{ij}K^{ij} \big) + \frac{6\sqrt{\g}}{4N^{5}}A_{0}^{5}\big( K\dot{K} - K_{ij}\dot{K}^{ij}\big)\,.
\een
These three terms cancel up to a boundary term after performing an integration by parts on the last term.  We again see that the Horndeski like tuning of the action depends on the fact that non-minimal couplings with the curvature tensors contain the same operator components as the non-linear combination of covariant derivatives.  We will see in section \ref{ssec:ms} that this property motivates the utility of the epsilon tensor construction from section \ref{ssec:cvGNMC} as such operators are generated with opposing signs thanks to the epsilon tensor and thus they are cancelled out. 

\subsection{\label{ssec:ms} Covariantisation via minimal substitution} 
In the previous section we saw that the non-minimal coupling term in the vector-Horndeski Lagrangian, once written in terms of ADM variables, contained terms that exactly cancelled the problematic terms containing the derivative of the lapse introduced by naive covariantisation.  We can view this property as being inherited from using the Horndeski Lagrangian for the non-minimal covariantisation of the decoupled longitudinal mode.\\
In this section we take this analogy further where, motivated by the recent results concerning beyond Horndeski theories \cite{Gleyzes:2014dya}, \cite{Lin:2014jga}, \cite{Gleyzes:2014qga}, we investigate the possibility that the vector-Galileons can be covariantised by minimal substitution.  In order to do this we work with the fully expanded epsilon tensor construction given by equations (\ref{eq:vGsys1}) to (\ref{eq:vGsys2}),\footnote{In short, we construct $\mc{L}_{N}$ with 2N-2 space-time indices contracted over two epsilon tensors.} and minimally covariantise by substituting the derivatives for covariant derivatives.\\ 
We work with the same ansatz as before and again focus on searching for the existence of potentially unstable terms of the form $A_{0}\dot{N}$ and $\dot{A}_{0}N$.  We find that in the covariantisation of this construction the terms involving the derivative of the lapse are generated in an antisymmetric combination such that they automatically cancel without the use of an additional counter term.     

\subsubsection{\label{ssec} Minimally covariantised quartic vector Galileon} 
The minimally covariantised quartic vector Galileon can be written as\footnote{Where for convenience we have chosen the extra free parameter to be $c_{2}=-\frac{1}{2}$.}
\begin{align} \label{eq:L4ms}
	\mc{L}^{(4)}_{vG|ms} =& \sqrt{-g}A_{\sigma}A_{\lambda} g^{\sigma \lambda }\Big( \d^{\mu}A_{\mu}\d^{\nu}A_{\nu} - \d^{\mu}A_{\nu}\d^{\nu}A_{\mu} \Big) \notag \\ 
	&+ 2\sqrt{-g}A_{\mu}A_{\nu} g^{\mu \sigma}\Big( \d^{\rho}A_{\sigma}\d^{\nu}A_{\rho} - \d^{\nu}A_{\sigma}\d^{\rho}A_{\rho} \Big)\,,
\end{align} 
which is the derivative term studied in section \ref{sec:nmcvg4} combined with another derivative term stemming from the extra antisymmetric sum over the first two indices.\\
Working with the ADM formalism and with our ansatz we find that the second term contains
\bee
  	 2\sqrt{-g}A_{\mu}A_{\nu} g^{\mu \sigma}\Big( \d^{\rho}A_{\sigma}\d^{\nu}A_{\rho} - \d^{\nu}A_{\sigma}\d^{\rho}A_{\rho} \Big) \supseteq \frac{1}{2}\sqrt{\gam}(A_{0})^{4}\Big( \frac{\dot{K}}{N^4}\Big)\,,
\een
which cancels the contribution from the first term containing the derivative of the lapse given by equation (\ref{eq:vG4unstable}) without the need of an additional non-minimal counter term.\\
An efficient way to realize the cancellation with this ansatz is found by utilizing the antisymmetric structure of the Lagrangian.  Indeed, the antisymmetric properties of the epsilon tensors make it straight forward to realize that these terms cancel without the use of non-minimal couplings such as those discussed in the previous section\footnote{Here we chose the value of the free parameter to be $c_{2}=0$.}
\begin{align}
	\mc{L}^{(4)}_{vG} =& \sqrt{-g}\ve^{\m_{1} \mu_{3}\mu_{5} \l}\ve^{\n_{2} \nu_{4} \nu_{6}}_{\quad \quad\,\, \l} \,A_{\m_{1}}A_{\n_{2}}\, \big(A_{\mu_{3}\nu_{4}}A_{\mu_{5}\nu_{6}}\big)\,,\notag \\ 
	= & N\sqrt{\g}\,\ve^{0\,\mu_{3}\mu_{5} \l}\ve^{0\, \nu_{4} \nu_{6}}_{\quad \quad \l} \,(A_{0})^2\, A_{\mu_{3}\nu_{4}}A_{\mu_{5}\nu_{6}}\,, \notag \\
	= &N\sqrt{\g}\,\ve^{0\,m_{3} m_{5} l}\ve^{0\,n_{4} n_{6}}_{\quad \quad l}\,(A_{0})^2 \,A_{m_{3}n_{4}}A_{m_{5}n_{6}}\,, \notag \\ 
	=& N\sqrt{\g}\,\ve^{0\,m_{3} m_{5} l}\ve^{0\,n_{4} n_{6}}_{\quad \quad l}\,(A_{0})^4 \,\Gam^{0}_{m_{3}n_{4}} \Gam^{0}_{m_{5}n_{6}}\,, \notag \\
	=& \sqrt{\g}\,\ve^{0\,m_{3} m_{5} l}\ve^{0\,n_{4} n_{6}}_{\quad \quad l}\,(A_{0})^4 \,\frac{K_{m_{3}n_{4}}K_{m_{5}n_{6}}}{N}\,, \notag \\ 
\end{align}
where we have used the results of appendix \ref{sec:Chris} for the Christoffel symbols and latin indices denote spatial components.\\  
With our ansatz we recover an antisymmetric combination of the extrinsic curvature, $K_{ij}$.  Inspecting its definition given by equation (\ref{eq:K}) reveals that there are no terms of the form $\dot{A}_{0}N$, or $A_{0}\dot{N}$ which provides evidence that the special antisymmetric structure of the vector Galileons allows them to be consistently covariantised by minimal substitution. 

\subsubsection{Minimally covariantised quintic vector Galileon}
We find that the kind of cancellation of terms found above for the minimally covariantised quartic vector Galileon is possible for the quintic vector Galileon as well.  As for the quartic, it is rather more efficient to make use of the antisymmetric properties of the epsilon tensors\footnote{Here we chose the value for the free parameter to be $d_{3}=0$.} 
\begin{align}
	\mc{L}^{(5)}_{vG} = &\sqrt{-g}\ve^{\m_{1} \mu_{3}\mu_{5} \mu_{7}}\ve^{\n_{2} \nu_{4} \nu_{6}\nu_{8}} \,A_{\m_{1}}A_{\n_{2}}\, \big(A_{\mu_{3}\nu_{4}}A_{\mu_{5}\nu_{6}}A_{\mu_{7}\nu_{8}}\big)\,,\notag \\ 
	=& N\sqrt{\g}\,\ve^{0\,\mu_{3}\mu_{5} \mu_{7}}\ve^{0\, \nu_{4} \nu_{6}\nu_{8}}\, (A_{0})^2 A_{\mu_{3}\nu_{4}}A_{\mu_{5}\nu_{6}}A_{\mu_{7}\nu_{8}}\,, \notag \\
	=& N\sqrt{\g}\,\ve^{0 \,m_{3} m_{5} m_{7}}\ve^{0\, n_{4} n_{6} n_{8}}\,(A_{0})^2 \,A_{m_{3}n_{4}}A_{m_{5}n_{6}}A_{m_{7}n_{8}}\,,  \notag \\ 
	=& N\sqrt{\g}\, \ve^{0\,m_{3} m_{5} m_{7}}\ve^{0\, n_{4} n_{6} n_{8}}\,(A_{0})^5\, \Gam^{0}_{m_{3}n_{4}} \Gam^{0}_{m_{5}n_{6}} \Gam^{0}_{m_{7}n_{8}}\,,  \notag \\
	=& \sqrt{\g}\,\ve^{0 \,m_{3} m_{5} m_{7}}\ve^{0\, n_{4} n_{6} n_{8}}\,(A_{0})^5\, \frac{K_{m_{3}n_{4}}K_{m_{5}n_{6}}K_{m_{7}n_{8}}}{N^2}\,,  
\end{align}
which shows that the quintic vector Galileon exhibits the same property as was found for the quartic.  Specifically, we recover an antisymmetric combination of extrinsic curvature terms, $K_{ij}$, suggesting that it can also be minimally covariantised.
In order to see the cancellation in detail we expand out the Lagrangian to find
\begin{align}
	\mc{L}^{(5)}_{vG} :=& \mc{L}^{(5)}_{vG|A} + \mc{L}^{(5)}_{vG|B}\,,\notag \\
	\mc{L}^{(5)}_{vG|A}=&\,\sqrt{-g}A_{\mu} A_{\nu}g^{\mu \nu}g^{\r\s}g^{\l \t} g^{\ve \de} \Big(A_{\r \s}A_{\l \t}A_{\ve \de} - 3A_{\r\s}A_{\l\ve}A_{\t\de} +2A_{\r\de}A_{\s\l}A_{\ve\t}\Big)\,, \notag \\ 
	\mc{L}^{(5)}_{vG|B}=&-6\sqrt{-g} A_{\mu}A_{\nu}g^{\mu\r}g^{\nu\s}g^{\l\t}g^{\ve\de}\Big( A_{\r\s}A_{\l[\t}A_{\ve]\de}+ A_{\r[\t}A_{\ve]\l}A_{\s\de}\Big)\,. 
\end{align}
First note that only $A_{00} \supseteq \dot{A}_{0}$ and $\dot{N}$.  Therefore we only need to consider the factors of this term.  We find that the first term contributes
\bee
	\mc{L}^{(5)}_{vG|A} \supseteq\, 6 \sqrt{\g} N A_{0}^{2} g^{00}A_{00} \big( g^{00}g^{ij}g^{kl}- 2 g^{ij}g^{0k}g^{0l}\big)\big( A_{i[j}A_{k]l}\big)\,,
\een
which is exactly cancelled by the contribution from the second term
\bee
	\mc{L}^{(5)}_{vG|B} \supseteq \, -6 \sqrt{\g} N A_{0}^{2} g^{00}A_{00} \big( g^{00}g^{ij}g^{kl}- 2 g^{ij}g^{0k}g^{0l}\big)\big( A_{i[j}A_{k]l}\big)\,.
\een
We have focused our attention on the covariantisation of quartic and quintic vector Galileons as these are the only terms that come with counter terms in the vector Horndeski system.  We have found evidence that, as with the recently proposed $G^3$ beyond Horndeski theories, the vector Galileons can be consistently covariantised by minimal substitution.

\subsubsection{\label{sec:eff} The effect of switching on the spatial components of $A_{\mu}$} 
So far we have investigated the cancellation of dangerous terms for a special ansatz, albeit with a general metric.  We have seen that the antisymmetric property of the epsilon tensor guarantees a cancellation.  In this subsection we investigate what happens to this cancellation once we remove the restriction of our ansatz and switch on the spatial components of $A_{\mu}$.  We start by examining the structure of the quartic Lagrangian
\begin{align}
	\mc{L}=&\sqrt{-g}\ve^{\m_{1} \mu_{3}\mu_{5} \l}\ve^{\n_{2} \nu_{4} \nu_{6}}_{\quad \quad\,\, \l} \,A_{\mu_{1}}A_{\nu_{2}}\d_{(\mu_{3}}A_{\nu_{4})}\d_{(\mu_{5}}A_{\nu_{6})}\,, \notag \\
	=& \sqrt{-g}\ve^{\m_{1} \mu_{3}\mu_{5} \l}\ve^{\n_{2} \nu_{4} \nu_{6}}_{\quad \quad\,\, \l} \,A_{\mu_{1}}A_{\nu_{2}}\Big( \p_{(\mu_{3}}A_{\nu_{4})}\p_{(\mu_{5}}A_{\nu_{6})} -2\p_{(\mu_{3}}A_{\nu_{4})}\G^{\l}_{\mu_{5}\nu_{6}} A_{\l}+ \G^{\r}_{\mu_{3}\nu_{4}}\G^{\l}_{\mu_{5}\nu_{6}} A_{\r}A_{\l}\Big)\,, \notag \\
	\supseteq& \sqrt{-g}\ve^{\m_{1} \mu_{3}\mu_{5} \l}\ve^{\n_{2} \nu_{4} \nu_{6}}_{\quad \quad\,\, \l} \,A_{\mu_{1}}A_{\nu_{2}}\Big(\G^{\r}_{\mu_{3}\nu_{4}}A_{\r}-2\p_{(\mu_{3}}A_{\nu_{4})} \Big)\G^{\l}_{\mu_{5}\nu_{6}}A_{\l}\,.
\end{align}
First notice that the first term is just the ordinary quartic vector Galileon with commuting derivatives and thus we need only concentrate on the second and third terms.  Furthermore, since only $\Gamma^{\mu}_{00}$ contribute $\dot{N}$ terms, if we choose at least one of either $\mu_{1}$ or $\nu_{2}$ to be zero, or if the zero components are shared between the two Christoffel symbols, then we do not recover any of the terms we are focussing on.  Therefore, we need only consider
\begin{align}
	\mc{L}=&\sqrt{-g}\,\ve^{\m_{1} \mu_{3}\mu_{5} \l}\ve^{\n_{2} \nu_{4} \nu_{6}}_{\quad \quad\,\, \l} \,A_{\mu_{1}}A_{\nu_{2}}\d_{(\mu_{3}}A_{\nu_{4})}\d_{(\mu_{5}}A_{\nu_{6})}\,, \notag \\
	\supseteq& \,2\,\sqrt{-g}\,\ve^{m_{1}m_{3} 0 \l}\ve^{n_{2} n_{4}0}_{\quad \quad \l} A_{m_{1}}A_{n_{2}}A_{m_{3}n_{4}}A_{00}\,, \notag \\
	=& \,2\sqrt{-g}\,g^{00}\big\{ g^{m_{1}n_{2}}g^{m_{3}n_{4}}-g^{m_{1}n_{4}}g^{m_{3}n_{2}}\big\}A_{m_{1}}A_{n_{2}}A_{m_{3}n_{4}}A_{00}\,,\notag \\
	\supseteq& -2\sqrt{\g}\big(\g^{m_{1}n_{2}}\g^{m_{3}n_{4}}-\g^{m_{1}n_{4}}\g^{m_{3}n_{2}} \big) A_{m_{1}}A_{n_{2}}A_{m_{3}n_{4}}\Big( \frac{1}{N}\dot{A}_{0}+ \p_{0}\big( \frac{1}{N}\big)A_{0}\Big)\,.
\end{align}
Since, $A_{mn}= \p_{(m}A_{n)}-\G^{0}_{mn}A_{0}-\G^{i}_{mn}A_{i}\supseteq \frac{1}{N}(A_{0}-A_{i}N^{i})K_{mn}$, we find an obstruction to the cancellation due to terms of the form (ignoring the contributions from the shift, $N_{i}$) 
\bee
	\mc{L} \supseteq \frac{\sqrt{\g}}{N^2}A_{0}^2A_{i}A_{j}(\g^{ij}\dot{K}-\dot{K}^{ij})\,,
\een
which remain after integration by parts.\\
We find a cancellation for the form of the non-minimal coupling inspired by Horndeski,\footnote{See also \cite{Deffayet:2009mn} for the generalisation for scalar fields to D dimensions.} where using the results of appendix \ref{sec:UnstCurv} gives us
\begin{align}
	\sqrt{-g}A^2A_{\mu}A_{\nu}G^{\mu\nu}&\supseteq N\sqrt{\g}\Big(-\frac{1}{N^2}\Big)A_{0}^2A_{i}A_{j}\frac{1}{N}\big(\g^{ij}\dot{K}-\dot{K}^{ij} \big)\,,\notag \\
	&=-\frac{\sqrt{\g}}{N^2}A_{0}^2A_{i}A_{j}(\g^{ij}\dot{K}-\dot{K}^{ij})\,.
\end{align}
Allowing for vector fields to have nonzero spatial components prevents us from relying on the epsilon tensor to provide a cancellation and this reduces the applicability of our analysis for the vector Galileons to the choice of a special ansatz.\\  
Something similar happens for the beyond Horndeski theories where the cancellation of the terms involving the derivative of the lapse in the action can be seen when the scalar field is used to select a preferred frame in which it depends only on time \cite{Gleyzes:2014dya},\cite{Lin:2014jga},\cite{Gleyzes:2014qga}.  Moreover, since the minimally covariantised quartic Galileon has been shown to be ghost free in all frames \cite{Deffayet:2015qwa} it would appear that this failure of cancellation in the action for general frames is simply a complication rather than a pathology.\\
For the vector theory the cancellation was for a chosen ansatz, rather than a preferred frame, however both rely on the absence of spatial components which spoil the cancellation by increasing the exponent of the lapse relative to the exponent for either $\dot{\phi}$ for the scalars or $A_{0}$ for the vectors.  Given that the non-linear structure for the vectors seems to impart the same behavior as that for the scalars, it seems likely that our choice of ansatz could be reinterpreted as a gauge choice for a preferred foliation.\footnote{This could possibly be achieved by fixing $A_{\mu}$ to be parallel to the unit normal vector $\n_{\mu}$ defined in appendix \ref{sec:ADM}.}  However this might not be possible and, indeed, an understanding of the types of theory for which this `unitary gauge' analysis is applicable remains an open question.\\
Another way to see the apparent resolution of the pathology for the minimally covariantised scalar Galileons, is to focus on their field equations.  For example, minimally substituting covariant derivatives into the quartic Galileon leads to third order derivatives of the metric and of the field appearing in the equations of motion \cite{Deffayet:2009wt}.  However, the Bianchi identities can be used to find a second order constraint equation that allows one to replace the higher order derivatives to recover a second order system \cite{Deffayet:2015qwa}. That the vector Galileons share the same special cancellations as their scalar counterparts could suggest that a similar constraint exists for these theories as well.\\
In summary, we have provided some  circumstantial evidence for the existence of 
a `beyond Horndeski' version of the vector Galileon theory. If it exists,   such a theory corresponds
to  a vector-tensor  system that, in analogy with
the scalar counterpart of \cite{Gleyzes:2014dya}, is   free of ghosts  thanks  to secondary
constraints that avoid the propagation of additional dangerous degrees of freedom.  

\section{\label{Sec:CG-Higgs} Covariantisation of the Galileonic Higgs system} 
Vector Galileons are a self derivative extension of the Proca action and therefore explicitly break any gauge symmetry associated with the kinetic term $-\frac{1}{4}F_{\mu\nu}F^{\m\n}$.  Up until now we have focused on massive vectors associated with an explicitly broken abelian gauge invariance.  Here we will discuss an extension to the usual complex scalar Higgs mechanism for spontaneously generating these terms.\\
A consistent non-linear extension to the Higgs mechanism is of course interesting for its relevance to potential applications to particle physics and superconductivity\footnote{See \cite{Weinberg:1996kr} for an excellent discussion.} but is also motivated by the possibility that there might be pathologies in the cosmological phenomenology of the non-minimally covariantised vector Galileons. Indeed, in \cite{Tasinato:2014mia} it was shown that the non-linear derivative interactions, that lead to interesting cosmological applications, might also cause the theory to suffer from strong coupling issues around non-trivial backgrounds.  The hope is that the additional scalar Galileon inherited from the Higgs dynamics might alleviate these strong coupling issues and enhance the cosmological applicability of these models.  Moreover, possible connections
with the general scenario of Higgs inflation (see e.g., \cite{Bezrukov:2013fka} for a recent review) could also be developed.\\  
In \cite{Hull:2014bga} it was shown how to extend the Higgs mechanism with a Galileonic symmetry to generate the vector-Galileons spontaneously. Interestingly, this Galileonic Higgs theory recovers a bi-Galileon system in its decoupling limit which, given the existence of non-renormalisation theorems \cite{Luty:2003vm,Nicolis:2004qq}, could further improve the phenomenological attractiveness of this set of theories.
\subsection{A Higgs mechanism for vector Galileons}
The U(1) invariant Lagrangian for this system is given in flat space by
\bee
	\mc{L}_{G-Higgs} = -\frac{1}{4}F_{\m\n}F^{\m\n} -(\mc{D}\phi)(\mc{D}\phi)^* + \mc{L}_{\mathrm{(8)}} + \mc{L}_{\mathrm{(12)}} + \mc{L}_{\mathrm{(16)}} + V(\phi)\,,
\een
where $V(\phi)$ is the usual Higgs potential, $V(\phi) = - \mu^2 \phi \phi^* +\frac{\lambda}{2} (\phi \phi^*)^2$, the gauge covariant derivative is given by $\mc{D}_{\mu}\phi = \p_{\m}\phi - igA_{\mu}$ and $\mc{L}_{\mathrm{(8)}}$, $\mc{L}_{\mathrm{(12)}}$, and $\mc{L}_{\mathrm{(16)}}$, are constructed out of antisymmetric combinations of the following gauge invariant operators
\begin{align}
	L_{\mu\nu}&\equiv \frac12 \left
	[(\mc{D}_{\mu}\phi)^{*}(\mc{D}_{\nu}\phi)+(\mc{D}_{\nu}\phi)^{*}(\mc{D}_{\mu}\phi)\right]\,,
	 \\ P_{\mu\nu}&\equiv  \frac12 \left[\phi^*\mc{D}_{\mu}\mc{D}_{\nu}\phi +\phi\,\left(
	 \mc{D}_{\mu}\mc{D}_{\nu}\phi\right)^* \right]\,,\\
	  \,Q_{\mu\nu} &\equiv \frac{i}{2} \left[
	  \phi \left( \mc{D}_{\mu}\mc{D}_{\nu}\phi\right)^* 
	  -\phi^*\mc{D}_{\mu}\mc{D}_{\nu}\phi 
	  	  \right]\,. 
\end{align}
Under a $U(1)$ transformation the gauge covariant derivative $\mc{D}$, the complex scalar field $\phi$, and the vector field $A_{\mu}$ transform as
\bea
 \phi&\to&\phi\,e^{i \,\xi} \,,\\
 A_\mu&\to&A_\mu +\frac{1}{g}\,\partial_\mu \xi 
\,,\\
 \cD_\mu \phi &\to&e^{i \,\xi}\, \cD_\mu \phi \,,\\
  \cD_\mu \,  \cD_\mu\, \phi &\to&e^{i \,\xi}\, \cD_\mu \,  \cD_\nu\,\phi 
 \,.\eea 
We make use of the antisymmetry of the \textit{epsilon tensor} to construct our higher order operators 
\begin{align} 
	\mc{L}_{\mathrm{(8)}} &= \frac{1}{2!\, \Lambda^4}\,\ve^{\a\beta\mu_{1}\mu_{2}}\ve_{\a \beta\nu_{1}\nu_{2}}\,\left(\, \a_{(8)} L_{\mu_{1}}^{\,\nu_{1}}P_{\mu_{2}}^{\,\nu_{2}}+\beta_{(8)}L_{\mu_{1}}^{\,\nu_{1}}Q_{\mu_{2}}^{\,\nu_{2}}\,\right)\,,&\label{eq: GL1} \\ 
	\mc{L}_{\mathrm{(12)}} &= \frac{1}{\Lambda^8}\,\ve^{\a\mu_{1}\mu_{2}\mu_{3}}\ve_{\a\nu_{1}\nu_{2}\nu_{3}}\,\left(\, \a_{(12)} L_{\mu_{1}}^{\,\nu_{1}}P_{\mu_{2}}^{\,\nu_{2}}P_{\mu_{3}}^{\,\nu_{3}}+\beta_{(12)}L_{\mu_{1}}^{\,\nu_{1}}Q_{\mu_{2}}^{\,\nu_{2}}Q_{\mu_{3}}^{\,\nu_{3}}\,\right)\,,&\label{eq: GL2}\\
	\mc{L}_{\mathrm{(16)}} &=\frac{1}{\Lambda^{12}} \,\ve^{\mu_{1}\mu_{2}\mu_{3}\mu_{4}}\ve_{\nu_{1}\nu_{2}\nu_{3}\nu_{4}}\,\left(\, 
	\al_{(16)} L_{\mu_{1}}^{\,\nu_{1}}P_{\mu_{2}}^{\,\nu_{2}}P_{\mu_{3}}^{\,\nu_{3}}P_{\mu_{4}}^{\,\nu_{4}} +\beta_{(16)}L_{\mu_{1}}^{\,\nu_{1}}Q_{\mu_{2}}^{\,\nu_{2}}Q_{\mu_{3}}^{\,\nu_{3}}Q_{\mu_{4}}^{\,\nu_{4}}\,\right)\,.&\label{eq: GL3}
\end{align}
These operators represent a non-linear extension of the Higgs mechanism via derivative self interactions.  They are suppressed by the appropriate power of an energy scale $\Lambda$ and are factored by dimensionless parameters $\a_{(i)}$ and $\b_{(i)}$.  In addition, their form is similar to Galileons and hence the consistency of their covariantisation is non-trivial.\\
Note that if we decompose the field into its norm and phase, $\phi = \vphi e^{i\pi}$, where $\vphi$ and $\pi$ are two real fields, the gauge invariant operators, $L$, $P$ and $Q$ can be re-expressed as
\begin{align}
	L_{\mu\nu}&= \p_{\mu}\vf \p_{\nu}\vf + g^2 \vf^2 \hat{A}_{\mu}\hat{A}_{\nu}\,,\label{eq:L}\\
	 P_{\mu\nu}&= \vf \p_{\mu}\p_{\nu}\vf - g^2\vf^2\hat{A}_{\mu}\hat{A}_{\nu}\,,\label{eq:P} \\Q_{\mu\nu} \,
	  &= \frac{g}{2}\,[\p_{\mu}(\vf^2\hat{A}_{\nu})+\p_{\nu}(\vf^2\hat{A}_{\mu})]\,,\label{eq:Q}
\end{align}
where $\hat{A}_{\m}\equiv A_{\m}-\p_{\m}\pi$ is a gauge invariant combination.\\
Using these relations to expand out the operators given in equations (\ref{eq: GL1}), (\ref{eq: GL2}) and (\ref{eq: GL3}) gives us a mixed scalar-vector theory.  We then rely on spontaneous symmetry breaking to generate the vector Galileons.\\  
The phenomenon of spontaneous symmetry breaking relies on the Higgs achieving a non-zero vacuum expectation value, $v$.  In addition to this, the Higgs field develops non-trivial dynamics via fluctuations about the vacuum.  In order to understand this we expand the field about the vacuum, $v$ with a small perturbation $h$  
\be
\vphi := (v+\frac{h}{\sqrt{2}})\,.
\ee
With this definition of $\vphi$ the expressions for our operators $L_{\mu\nu}$, $P_{\mu\nu}$, $Q_{\mu\nu}$, and $V(\phi)$, become
\begin{align}
	L_{\mu\nu} \to& \,\frac{1}{2}\p_{\mu}h\p_{\nu}h + g^2\big( v+h/\sqrt{2} \big)^2\hat{A}_{\mu}\hat{A}_{\nu}\,,\\
	P_{\mu\nu} \to& \,\frac{\sqrt{2}}{2}\big( v+h/\sqrt{2}\big)\p_{\mu}h\p_{\nu}h - g^2 \big( v+h/\sqrt{2}\big)^2\hat{A}_{\mu}\hat{A}_{\nu}\,,\\
	Q_{\mu\nu} \to& \,g \sqrt{2} \big( v+h/\2 \big)[ \p_{(\mu}h\hat{A}_{\nu)}+ \frac{\2}{2}\big( v+h/\2 \big)\hat{A}_{\mu\nu} ]\,,\\
	V(\phi) \to& -\frac{\lam}{2} v^4 + \lam v^2 h^2 +\frac{\2}{2}v\lam h^3 + \frac{\lam}{8}h^4\,.
\end{align}
For purpose of demonstration, we restrict our discussion to a subset of operators from $\mc{L}_{G-Higgs}$ and recover the quartic vector Galileon by expanding around the background of the Higgs' \textit{vev}
\begin{flalign}\label{eq:Lg-higgs}
	\mc{L}_{\hat{A},h} =& -\frac{1}{4} F_{\mu \nu} F^{\mu \nu} - \big(m_{A}^2-\sqrt{2}g m_{A}h-\frac{1}{2}g^2h^2\big)\hat{A}^2 \nonumber& \\ &- \frac{1}{2}(\p h)^2 +\frac{\lam}{2}v^4 - \frac{1}{2}m_{h}^2 h^2 -\frac{\sqrt{\lam}m_{h}}{2}h^3 -\frac{\lam}{8}h^4 \nonumber& \\ &+ \frac{1}{\Lambda^{8}} \ve^{\m_{1} \mu_{3}\mu_{5} \l}\ve^{\n_{2} \nu_{4} \nu_{6}}_{\quad \quad\,\, \l} \big( v+\frac{\2}{2}h \big)^{2} \Big[ \frac{1}{2} h_{\m_1} h_{\n_2} + g^2 \big( v + \frac{\2}{2} h \big)^2 \hat{A}_{\m_1} \hat{A}_{\n_2} \Big]&\nonumber \\  &\cdot \Big\{ \beta_{(12)}\,(\2 g )^{2}\Big[ h_{(\m_{3}} \hat{A}_{\n_{4})} + \frac{2}{\2} \big( v + \frac{\2}{2} h \big) \hat{A}_{\m_{3}\n_{4}} \Big] \dots \nonumber& \\ &\qquad \qquad \qquad \qquad \qquad \quad \dots \Big[ h_{(\m_{5}} \hat{A}_{\n_{6})} + \frac{2}{\2} \big( v + \frac{\2}{2} h \big) \hat{A}_{\m_{5} \, \n_{6}} \Big] \Big\}\,,&
\end{flalign}
with
\bea
m_A&\equiv& g \,v\,,
\\
m_h&\equiv&\sqrt{2\,\lambda}\,v\,.
\eea
The expansion about the Higgs's \textit{vev} and the resulting Lagrangian, $\mc{L}_{\hat{A},h}$ in equation (\ref{eq:Lg-higgs}), describes a fully $U(1)$ invariant theory of four degrees of freedom: two scalars $\pi,h$ and a massless vector $\hat{A}_{\mu}$.  However, in general, we must go to the unitary gauge ($\pi =0$) to reveal the true physical degrees of freedom.  Using this gauge we find that our theory describes a physical system with two interacting fields: a scalar field $h$, representing the Higgs, interacting with a massive vector field $\hat{A}_{\mu}$.  Furthermore, we see that indeed not only does the \textit{vev} of the Higgs produce a mass for the vector boson but in this restricted case we also recover a higher dimensional operator that resembles the quartic vector Galileon that was first studied in \cite{Tasinato:2014eka,Heisenberg:2014rta}.\footnote{That is, up to an additional free parameter and a possible intergration by parts.}  In order to connect with those results, we expand out the terms in equation (\ref{eq:Lg-higgs}) which are both $O(h^0)$ and $O(\hat{A})$ and above
\bee
	\mc{L}_{\hat{A}} = -\frac{1}{4} F_{\mu \nu} F^{\mu \nu} -m_{A}^2 \hat{A}^2 +  \tilde{\beta}_{(12)} \ve^{\m_{1} \mu_{3}\mu_{5} \l}\ve^{\n_{2} \nu_{4} \nu_{6}}_{\quad \quad\,\, \l}\,\hat{A}_{\m_1}\hat{A}_{\n_2}\hat{A}_{\m_{3}\n_{4}}\hat{A}_{\m_{5} \, \n_{6}}\,,
\een
where
\bee
	\tilde{\beta}_{(12)} \equiv \frac{\beta_{(12)}g^{4}v^{6}}{\Lambda^{8}}\,.
\een
Which shows we have indeed recovered the quartic vector Galileon [see equation (\ref{eq:vGsys4})].\\ 
In \cite{Hull:2014bga} it was shown that there exists a limit in which this strong coupling scale remains fixed but the transversal vector degrees of freedom are decoupled.  Furthermore, the interactions between the longitudinal modes and the Higgs form a separate bi-Galileon system.  In this decoupling limit we calculate the strong coupling scale of the theory to be
\begin{equation}
	\Lambda_g \sim \Big(\frac{\beta_{(12)}g^{4}v^{6}}{m_{A}^4}\Big)^{1/8}.
\end{equation}

We have reviewed the mechanism discussed in \cite{Hull:2014bga} that generates the vector Galileons via spontaneous symmetry breaking.  Since we are interested in cosmological applications, it is important to covariantise the theory.  In the next section we will discuss the different covariantisation schemes that can be applied and their corresponding issues. 

\subsection{Covariantisation of the system}
In this section we consider the effects of covariantising this system on general space-times.  Since we have constructed the system using the same non-linear  structure as that of the Galileons, it would appear that after minimal substitution with covariant derivatives the terms up to cubic order remain ghost free whereas the terms of quartic and higher order could introduce ghosts and therefore need careful consideration.\\
It has been proven in \cite{Deffayet:2015qwa} that the pure scalar sector of $\mc{L}_{\mathrm{(12)}}$, which corresponds to a quartic Galileon, can be consistently covariantised by minimal substitution.  Furthermore, in section \ref{sec:cvg} we found evidence that this is also a consistent way to write a covariant theory for the vector Galileons and hence the pure vector sector of $\mc{L}_{\mathrm{(12)}}$.  We are able to reproduce these sectors if we re-write the above expressions in terms of new covariantised operators constructed from replacing partial derivatives with covariant derivatives: $ \tL_{\m\n}:=L_{\m\n}|_{\p\to\d} \equiv L_{\m\n}$, $\tP:=P_{\m\n}|_{\p\to\d}$, and $ \tQ_{\m\n}:=Q_{\m\n}|_{\p\to\d}$.  However, it is not immediately clear whether such a process interferes with the $U(1)$ gauge invariance of the operators.  Indeed, for terms with only one partial derivative or an undifferentiated gauge vector, there will be no change and therefore as the notation suggests, there is no problem with the $L$ operator.  However, the two remaining operators depend upon $\phi^*\mc{D}_{\mu}\mc{D}_{\nu}\phi$ and therefore we should check the effect of covariantising the partial derivatives 
\bee
	\phi^*\mc{D}_{\mu}\mc{D}_{\nu}\phi = \phi^* \d_{\m}\p_{\n}\phi - i\phi^* \phi\d_{\m}A_{\n} -i\phi^* A_{\n}\p_{\m}\phi -i \phi^*A_{\m}\p_{\n}\phi -A_{\m}A_{\n}\phi^*\phi\,.
\een
Although we find two additional terms due to the covariantisation of the partial derivatives, this does not spoil the gauge invariance as for $\phi \to\phi\,e^{i \,\xi}$ and $A_\mu \to A_\mu + \partial_\mu \xi$ we find  
\bee
	-\Gamma^{\l}_{\m\n} \Big( \p_{\l}\phi -iA_{\l}\phi \Big) \to -e^{i\,\xi}\Gamma^{\l}_{\m\n} \Big( \p_{\l}\phi -iA_{\l}\phi \Big)\,, 
\een 
where the multiplicative factor of $e^{i\,\xi}$ is cancelled by the contribution coming from $\phi^*\to e^{-i\,\xi}\phi^*$.\\
We have established that generalizing our operators for curved spacetimes does not interfere with their $U(1)$ gauge invariance.  However, it is not yet clear how we should approach the mixing terms that would also be generated as in this case we lose the utility of choosing a special ansatz.  This is also true for the decoupling limit of the theory, which is a system of bi-Galileons, whose minimal covariantisation, if its consistency were to be established, would resemble a multi-field generalization of the beyond Horndeski theory.  \\
In the next section, we investigate a safer way to remove the problematic terms involving the derivative of the lapse by introducing non-minimal couplings.  
  
\subsection{$U(1)$ invariant non-minimal couplings}
In this section we investigate the form of the possible non-minimal couplings we could add to the theory.  These should be compatible with $U(1)$ invariance and valid in all frames.\\
In section \ref{sec:cvg} we presented a consistent non-minimal covariantisation for vector Galileons.  This result together with the Horndeski system \cite{Horndeski:1974wa} provides a consistent non-minimal covariantisation for both the pure scalar and vector sectors of the Galileonic Higgs.  However, the consistency and effectiveness of the non-minimal counter terms for the mixed scalar vector sector still needs to be addressed.  Subsequently, we comment on the correspondence of the decoupling limit of this theory to the multi-field generalization of the Horndeski system proposed by \cite{Ohashi:2015fma}.\\
For inspiration we start with analyzing the scalar sector and focus on the non-minimal covariantisation of $\mc{L}_{(12)}^{\alpha}$.
\subsubsection{Non-mimimal coupling for $\mc{L}_{(12)}$}
First we set $g=1$, define $\vphi \hat{A}_{\mu}:= A_{\mu}$ and expand out the terms from equations (\ref{eq:L}), (\ref{eq:P}), and (\ref{eq:Q}).\\
Notice that the antisymmetry of the epsilon tensors guarantees that we cannot have more than two vector fields.  Therefore we can write $\mc{L}_{(12)}^{\alpha}$ as
\begin{align}
	\mc{L}_{(12)} \supseteq \sqrt{-g} \,\frac{\al_{(12)}}{\Lambda^8} \ve^{\m_1 \m_3 \m_5 \l}\ve^{\n_2 \n_4 \n_6}_{\qquad \,\,\l} \Big(&\underbrace{(\vphi_{\m_1} \vphi_{\n_2}  +  A_{\m_1}A_{\n_2})\vphi^2 \vphi_{\m_3 \n_4} \vphi_{\m_5 \n_6}}_{(A)} \notag \\ 
	-&\underbrace{ \vphi \vphi_{\m_1}\vphi_{\n_2}(A_{\m_3} A_{\n_4} \vphi_{\m_5 \n_6} + \vphi_{\m_3 \n_4} A_{\m_5} A_{\n_{6}})}_{(B)}\Big)\,. 
\end{align}
The contribution to the equation of motion for $\vphi$ from term (B) does not produce any higher order derivatives on the metric, however the contribution from term (A) does.  In order to find a consistent non-minimal covariantisation we follow the method demonstrated by \cite{Deffayet:2009mn} and find a term that mixes the derivatives of the scalar with the curvature tensor.  This type of non-minimal coupling to gravity, called \textit{kinetic gravity braiding}, introduces interesting cosmological phenomenology to our model.  However, we must also guarantee that our action remains U(1) invariant.  We should therefore gauge covariantise the derivatives of the scalar and thus introduce a non-minimal coupling between the vectors and the curvature tensor.\\
By taking a variation with respect to the scalar field, a higher derivative term is derived from (A) and can be expressed as 
	\bee
		\sqrt{-g} \,\frac{\al_{(12)}}{2\Lambda^8}\ve^{\m_1 \m_3 \m_5 \l}\ve^{\n_2 \n_4 \n_6}_{\qquad \,\,\l}\delta \vphi \vphi^{2}(\vphi_{\m_1} \vphi_{\n_2}  + A_{\m_1}A_{\n_2})\d^{\lambda}\vphi R_{\n_{4}\n_{6}\m_{3}\m_{5};\lambda}\,.
	\een
In order to remove this we are required to have the following additional term in the Lagrangian 
\bee
\sqrt{-g}\,\frac{\al_{(12)}}{4\Lambda^8}\ve^{\m_1 \m_3 \m_5 \l}\ve^{\n_2 \n_4 \n_6}_{\qquad \,\,\l} \vphi^2 \d^{\lambda}\vphi \d_{\lambda}\vphi(\vphi_{\m_1} \vphi_{\n_2}  + A_{\m_{1}} A_{\n_{2}}) R_{\n_{4}\n_{6}\m_{3}\m_{5}}\,,
\een 
which cures both the pure scalar sector and the scalar vector cross terms.  However this is not U(1) invariant.  We can make this U(1) invariant by `gauge covariantising' the covariant derivatives: $\d_{\lambda} \vphi \to \mc{D}_{\lambda}\phi \equiv \d_{\lambda}\phi - i A_{\lambda}\phi$.  We can then write this in terms of the operators in our theory as
	\bee
		\mc{L}_{(12)}^{\alpha NMC} = \sqrt{-g}\,\frac{\al_{(12)}}{4\Lambda^8}\ve^{\m_1 \m_3 \m_5 \l}\ve^{\n_2 \n_4 \n_6}_{\qquad \,\,\l} \phi^*\phi L L_{\m_{1}\n_{2}} R_{\m_{3}\m_{5}\n_{4}\n_{6}}\,.
	\een 
In order to get a better intuitive picture about this term we expand out the sum
	\begin{align}
		\ve^{\m_1 \m_3 \m_5 \l}\ve^{\n_2 \n_4 \n_6}_{\qquad \,\,\l}  L_{\m_{1}\n_{2}} R_{\m_{3}\m_{5}\n_{4}\n_{6}} &= L R - L R^{\rho \sigma}_{\,\,\,\,\,\,\rho \sigma} + L_{\mu\nu} \big ( R^{\rho\nu\mu}_{\,\,\,\,\,\,\,\,\rho} -R^{\nu\rho\mu}_{\,\,\,\,\,\,\,\,\,\rho}+R^{\nu\rho \,\,\mu}_{\,\,\,\,\,\rho} -R^{\rho \nu \,\,\mu}_{\,\,\,\,\,\rho} \big)\,, \notag \\ 
		&= 2L R-4 L_{\mu\nu} R^{\mu\nu}\,, \notag \\ 
		&= 4L_{\mu\nu} \big( \frac{1}{2}g^{\mu\nu}R-R^{\mu\nu}\big)\,, \notag \\ 
		&= -4L_{\mu\nu}G^{\mu\nu}\,.
	\end{align}
We can now write our non-minimal coupling term as
	\bee \label{eq:GHNMC}
		\mc{L}_{(12)}^{\alpha NMC} = -\sqrt{-g}\frac{\al_{(12)}}{\Lambda^8} \phi^*\phi L L_{\mu\nu}G^{\mu\nu}\,.
	\een
Notice that we now have extra cross terms coming from the need to make our counter term $U(1)$ gauge invariant.  Although this might seem at first to be pathological as it introduces new higher derivative terms, we will see contrary to this, that a solution can be found in the form of a unique theory.\\  
We now go through the same process to find the counter term for the vector sector.  Again we concentrate on the quartic and expand the term factored by $\beta_{(12)}$ which contains the quartic vector Galileon plus mixed scalar and vector terms
\bee
	\mc{L}_{(12)} \supseteq \sqrt{-g} \,\frac{\beta_{(12)}}{\Lambda^8} \ve^{\m_1 \m_3 \m_5 \l}\ve^{\n_2 \n_4 \n_6}_{\qquad \,\,\l} 
	(\vphi_{\m_1} \vphi_{\n_2}  +  A_{\m_1}A_{\n_2})\vphi^2 A_{\m_3 \n_4} A_{\m_5 \n_6}\,. 
\een
We know from section \ref{sec:eff} that for this particular form of the vector sector we require a counter term of the form
\bee
	 \sim -\sqrt{-g}A^{2}\big(\p_{\mu}\vphi\p_{\nu}\vphi+A_{\mu}A_{\nu}\big)G^{\mu\nu}\,,
\een
which cures both the pure vector sector and the scalar vector cross terms.  To ensure our counter term is $U(1)$ invariant, we form the trace of the gauge invariant operator, $\mathrm{tr}L_{\m\n} = L$, out of the $A^2$ factor.  Thus we end up with the same form of counter term as we found for the scalar sector
 	\bee
		\mc{L}_{(12)}^{\beta NMC} = -\sqrt{-g}\frac{\beta_{(12)}}{\Lambda^8} \phi^*\phi L L_{\mu\nu}G^{\mu\nu}\,.
	\een
Here we conclude that the combination of both covariantisation and $U(1)$ gauge invariance has ensured that we recover the same form for the non-minimal coupling for both branches of the quartic Galileonic Higgs.  	
\subsubsection{Cross terms}
We have found that $U(1)$ gauge invariance requires the counter terms constructed for both the scalar and vector sectors of the quartic to be identical.  In order to be satisfied that this is the correct choice of counter term we must also check whether we recover the right form for the mixed vector scalar terms.  We begin by expanding out the gauge invariant operators in $\mc{L}_{(12)}^{\al}+\mc{L}_{(12)}^{\beta} + \mc{L}_{(12)}^{ \alpha NMC}$ and examining the cross terms
\begin{align}
	\mc{L}_{(12)}^{\al}+ \mc{L}_{(12)}^{\beta}+\mc{L}_{(12)}^{\alpha NMC} &\supseteq \sqrt{-g} \,\frac{\al_{(12)}}{\Lambda^8} \ve^{\m_1 \m_3 \m_5 \l}\ve^{\n_2 \n_4 \n_6}_{\qquad \,\,\l} (\vphi_{\m_1} \vphi_{\n_2}  +  A_{\m_1}A_{\n_2})\vphi^2 \vphi_{\m_3 \n_4} \vphi_{\m_5 \n_6}\notag \\ 
	&+\sqrt{-g} \,\frac{\beta_{(12)}}{\Lambda^8} \ve^{\m_1 \m_3 \m_5 \l}\ve^{\n_2 \n_4 \n_6}_{\qquad \,\,\l} 
	(\vphi_{\m_1} \vphi_{\n_2}  +  A_{\m_1}A_{\n_2})\vphi^2 A_{\m_3 \n_4} A_{\m_5 \n_6}\notag \\
	&-\sqrt{-g}\frac{\al_{(12)}}{\Lambda^8} \phi^*\phi (\underbrace{\vphi_{\l}\vphi^{\l}}_{I} + \underbrace{A_{\l}A^{\l}}_{II})( \vphi_{\mu}\vphi_{\nu}+ A_{\m}A_{\n})G^{\mu\nu} \,. 
\end{align}
The non-minimal coupling factored by the term labelled $I$ was necessary for the consistency of $\mc{L}_{(12)}^{\al}$ but $U(1)$ gauge invariance forced us to include the term factored by $II$.  This extra term, however, turns out to be exactly the form of non-minimal coupling necessary for the consistency of $\mc{L}_{(12)}^{\beta}$.  This suggests that in order to generate the correct combination of terms for the pure vector and scalar sectors as well as the mixing terms we simply need to factorize each term with the appropriate dimensionless parameter.  On the other hand, these parameters cannot  be independent as this would be inconsistent with $U(1)$ gauge invariance. Therefore a consistent non-minimal covariantisation of the quartic Galileonic Higgs is with $\al_{(12)} = \beta_{(12)}=\g_{(12)}$
\bee
	\mc{L}_{(12)} = \mc{L}_{(12)}^{\al}|_{\al=\g} + \mc{L}_{(12)}^{\beta}|_{\b=\g} + \mc{L}_{(12)}^{\g NMC}\,,
\een   
where $\mc{L}_{(12)}^{\g NMC}$ is that of equation (\ref{eq:GHNMC}) with $\al_{(12)}$ substituted with a new dimensionless parameter, $\g_{(12)}$.\\
In the process of constructing a non-minimal covariantisation of the quartic Galileonic Higgs we have found that, although around Minkowski we can have two separate sectors of the theory parametrized by $\al_{(12)}$ and $\beta_{(12)}$, on generally curved space-times consistency with $U(1)$ gauge invariance requires them to be equal.  Furthermore, we find that the form of the unique counter term simultaneously compatible with both generally curved space-times and gauge invariance is closely related to that suggested for the generalized multi-field quartic by \cite{Ohashi:2015fma}.  Indeed, in an appropriate decoupling limit, we find that this counter term would exactly resemble that for the covariantised quartic bi-Galileon.  This is consistent with the covariantisation of the decoupling limit of our theory as around Minkowski space we find in such a limit that the Galileonic Higgs reduces to a bi-Galileon system.  

\section{ \label{Sec:Conc} Discussion}

Vector Galileons  are ghost-free
 systems containing derivative self-interactions of vector fields, 
 that break gauge symmetry, and that  can have interesting cosmological consequences thanks
 to their relation with the scalar Galileons.   
  In this paper, we have made use of    an ADM approach
to    carefully reconsider consistent covariant couplings
of vector Galileons with gravity that were first  introduced and analyzed  in  \cite{Tasinato:2014eka,Heisenberg:2014rta} by working
 in analogy with the scalar Galileon counterparts. We started by discussing non-minimal couplings of vector Galileons
 with gravity, studying the conditions to avoid the propagation of ghosts. We then continued with a discussion on minimal couplings
 with gravity, with the purpose of investigating  possible  vector-tensor counterparts to the `beyond Horndeski' scalar-tensor theories of \cite{Gleyzes:2014dya}. Our analysis indeed led us to speculate about, and provide some circumstantial evidence for, the existence of a general  `beyond vector Horndeski' vector tensor theory, in which secondary constraints avoid the propagation of additional ghostly  degrees of freedom.  We leave for future work a detailed  analysis of this theory.\\
In the second part of this work, we studied the covariantisation of a Higgs system that leads to the vector Galileon theories, after a spontaneous breaking of the gauge symmetry \cite{Hull:2014bga}.  Such an extension to the Higgs model requires higher order derivative self-interactions, that are  ghost-free. Its  covariantisation can be pursued along the lines and  with the same techniques 
  discussed in the first part of the work. The quartic system induces a unique, non-minimal, gauge invariant derivative coupling of the Higgs 
   scalar with curvature that can be related to the counter term suggested by \cite{Ohashi:2015fma}.  Such couplings can be extended to the non-Abelian case, developing the arguments introduced in \cite{Hull:2014bga} for the model in Minkowski space. Unfortunately, following the same procedure for the quintic leads to difficulties.  The U(1) gauge covariantisation of the counter terms reintroduces higher derivatives of the metric thus spoiling the consistency of the theory.  This suggests that, for the Galileonic Higgs system, simultaneous gauge and gravitational covariantisation is not compatible for the quintic theory.\\
Studies of similar systems of gauged Galileons and their covariantisation were carried out in \cite{Zhou:2011ix,Goon:2012mu}.  In particular, \cite{Goon:2012mu} were able to generate the terms from \cite{Zhou:2011ix} from a braneworld scenario by gauging the isometries in the bulk.  It would be worthwhile investigating whether a similar result can be found for the model we have presented in this work, as such a scenario might provide insight towards a generalized gauge Higgs unification.\\
It would be interesting to study whether non-minimal derivative couplings of the Higgs with gravity -- allowed by the gauge symmetries -- can have some consequences for the general scenario of Higgs inflation (see e.g., \cite{Bezrukov:2013fka}). In section \ref{Sec:CG-Higgs} we calculated the strong coupling scale for the quartic theory which naively suggests that the theory would be strongly coupled in inflationary scenarios.  However, the theory might still be trustworthy at this energy regime as the determination of the strong coupling scale for theories with higher derivatives depends on the background.  We leave this question, as well as other possible applications to cosmology, for future work.

\acknowledgments
MH would like to thank Daniel Goddard, David Langlois, Shinji Mukohyama and Ryo Namba for useful discussions.  MH is supported by a U.K. Science and Technology Facilities Council (STFC) research studentship.  GT is supported by an STFC Advanced Fellowship ST/H005498/1. KK is supported by the STFC through the consolidated grant ST/K00090X/1, and the European Research Council through grant 646702 (CosTesGrav). 

\appendix

\section{ADM-Decomposition}\label{sec:ADM}

\subsection{ADM-Decomposition of the metric}  Given a four-dimensional space-time $\mc{V}$, we may introduce a scalar field $t(x^{\alpha})$ such that $t=const$ defines a family of non-intersecting, spacelike, three surfaces $\Sigma_{t}$.  This allows us to introduce a foliation of the four dimensional space-time such that the metric, $g_{\mu\nu}$, can be decomposed in terms of components normal and tangent to the three surfaces $\Sigma_{t}$
\bee
	g_{\mu\nu} = -\big( N^2-N_iN^i\big)dt^2 +2 N_i dt dx^i + \gamma_{ij}dx^i dx^j\,,
\een
where $N$ is called the \emph{lapse} and $N_i$ is the \emph{shift}.\\  We write the unit normal to the three surface as $n^{\mu}=\big( -\frac{1}{N},\frac{N^i}{N}\big)$ with corresponding one-form $n_{\mu} = \big(-N,\vec{0} \big)$.  Furthermore, equivalently we may write: $g_{00} = -(N^2 - N_i N^i )$, $g_{0i} = N_i$, and $g_{ij}=\gamma_{ij}$.  The associated four dimensional inverse metric's components may be written as: $g^{00}=-\frac{1}{N^2}$, $g^{0i} = \frac{N^i}{N^2}$, and $g^{ij}=\gamma^{ij} - \frac{N^i N^j}{N^2}$.  The corresponding metric determinants are associated by: $\sqrt{-g} = N \sqrt{\gamma}$.\\
We define the extrinsic curvature of the spatial slices $t=constant$ as
\bee \label{eq:K}
	K_{ij}\equiv - \d_{i}n_{j} = \Gamma^{\mu}_{ij}n_{\mu} = -N \Gamma^{0}_{ij} = \frac{1}{2N}\Big( D_{i}N_{j}+D_{j}N_{i}- \dot{\gam}_{ij} \Big)\,.
\een
  
\subsection{Formula for Christoffel symbols} \label{sec:Chris}
The four dimensional connection is given by, $\G^{\mu}_{\nu\rho} = \frac{1}{2}g^{\mu \sigma}(g_{\nu \sigma, \rho}+ g_{\rho\sigma, \nu} - g_{\nu \rho,\sigma} )$.  We will find it useful to first collect the results for all the connection components in one place
\begin{align}
	\Gamma_{ij0}=\Gamma_{i0j} &= -NK_{ij} + D_{j}N_{i}\,, \\ 
	\Gamma_{ijk} &= \tG_{ijk}\,,\\ 
	\Gam^{0}_{00} &= \frac{1}{N}\big( \dot{N} + N^{i}\p_{i} N-N^{i} N^{j} K_{ij}\big)\,, \\ 
	\Gam^{0}_{0i} = \G^{0}_{i0} &=  \frac{1}{N}\big(\p_{i}N -N^{j}K_{ij}\big)\,,\\ 
	\Gam^{i}_{0j} = \G^{i}_{j0} &= -\frac{N^{i}\p_{j}N}{N}-N\Big( \g^{ik}-\frac{N^{i}N^{k}}{N^2}\Big)K_{kj} + D_{j}N^{i}\,,\\ 
	\Gam^{0}_{ij} &=-\frac{1}{N}K_{ij}\,,\\ 
	\G^{i}_{jk}&=\tG^{i}_{jk} + \frac{N^{i}}{N}K_{jk}\,,\\ 
	\G^{i}_{00} &=- \frac{\dot{N}N^{i}}{N} + \g^{ij}\dot{N}_{j}+ \frac{1}{2N^{2}}N^{i}N_{k}N_{l}\dot{\g}^{kl}\notag \\
	&+ \frac{1}{2}\Big( \g^{ij}-\frac{N^{i}N^{j}}{N^2}\Big)\p_{j}(N^2-N_{k}N^{k})\,.
\end{align}

\subsection{Unstable terms from the curvature tensor components} \label{sec:UnstCurv}
In the ADM formalism the Ricci scalar, $R$ is given by
\begin{align}
	R= &g^{\mu\nu}g^{\al\beta}R_{\mu\al\nu\beta} = \gam^{ik}\gam^{jl}R_{ijkl}-2n^{\mu}n^{\nu}\gam^{ij}R_{\mu i \nu j}\,,\notag \\
	=&\tilde{R} +K^{ij}K_{ij}+K^{i}_{i}K^{j}_{j}-\frac{2}{N}\dot{K}^{i}_{i} + 2\frac{N^{j}}{N}D_{j}K^{i}_{i}-\frac{2}{N}D^2N\,.
\end{align}
We see that it contains
\bee
	R \supseteq -\frac{2}{N}\dot{K}^{i}_{i}\,,
\een
which from the definition of $K_{ij}$ from equation (\ref{eq:K}) must contain $\dot{N}$.\\
The spatial components of the Ricci curvature tensor contains
\bee
	R^{ij} \supseteq \frac{N^{i}N^{j}\dot{K}}{N^{3}}-\frac{\g^{ik}\g^{jl}}{N}\dot{K}_{kl}\,.
\een
With these we find that the spatial components of the Einstein tensor, $G^{ij}$ therefore contains
\bee
	G^{ij} \supseteq \frac{1}{N}\Big( \g^{ij}\dot{K} - \g^{ik}\g^{jl}\dot{K}_{kl}\Big)\,.
\een
Which contains time derivatives of the lapse, $N$.


%

\end{document}